# Experimental Realization of Topological Surface Acoustic Wave Resonances under Surface Liquid Loading for Micro-volume sample Sensings

Bowei Wu [1, 2], Tingfeng Ma [1, 2, *], Hanbang Deng [1, 2], Shuanghuizhi Li [1, 2]

[1] Zhejiang-Italy Joint Lab for Smart Materials and Advanced Structures, School of Mechanics and Engineering Science, Ningbo University, Ningbo 315211, China.

[2] Key Laboratory of Impact and Safety Engineering, Ministry of Education, Ningbo University, Ningbo 315211, China

* Author to whom correspondence should be addressed: matingfeng@nbu.edu.cn

## Abstract

Surface acoustic wave (SAW) sensors, owing to their high operating frequencies, compatibility with planar microfabrication, and exceptional sensitivity to surface perturbations, are widely used in lab-on-a-chip and biomedical diagnostic applications. However, conventional SAW sensors operating in liquid environments suffer from substantial energy dissipation, which markedly reduces their quality factors (Q factors) and limits sensing reliability. Here, we design and fabricate a topological SAW resonant sensor for surface liquid loading by exploiting interference coupling between topological interface states and resonant cavities. The band structures are validated using a semi-analytical plane wave expansion–finite element method. The device requires only 0.04 μL of sample and enables concentration sensing of NaCl and glucose solutions, with sensitivities of up to 190 kHz/% for NaCl and 101.8 kHz/% for glucose, respectively, and a concentration resolution of 0.01%. This work offers a potential technological route towards high-performance SAW biomedical chips for micro-volume sample analysis.

## 1. Introduction

Surface acoustic wave (SAW) devices are key components in wireless communications[1] and high-frequency sensing[2,3]. Their acoustic energy is confined within approximately one wavelength of the surface of a piezoelectric substrate, rendering them intrinsically sensitive to perturbations in surface mass loading, viscosity, and electrical conductivity. They have therefore been widely used in chemical and biological sensing[4–6]. However, SAW propagation is accompanied by substantial energy dissipation[7], which hinders highly localized energy transport and markedly degrades the quality factor and overall sensing performance of the device.

The introduction of phononic crystals (PnCs) has opened new avenues for acoustic-wave manipulation. Incorporating defect states into PnCs can enhance acoustic-wave transport and localization[8,9]. Although defect-state PnC sensors can strengthen interactions between acoustic waves and target media, they are highly sensitive to fabrication errors. Minor geometric deviations and processing defects can induce pronounced backscattering and energy dissipation, causing resonance-frequency shifts and reduced quality factors, thereby compromising sensing reliability.

Topological phononic crystals provide a promising strategy for suppressing scattering losses and improving device robustness[10,11]. Breaking the spatial inversion symmetry of a periodic lattice can generate valley-Hall topological interface states at the boundary between distinct topological phases. These interface states confine acoustic energy near predefined propagation paths. They can also suppress backscattering induced by weak disorder and local defects[12–15]. Previous studies have demonstrated robust transport in topological SAW waveguides on semi-infinite substrates[16–18]. However, surface liquid loading alters the boundary conditions, band structure, and modal energy distribution of the device. Its effects on topological resonant behavior remain unexplored experimentally. Under surface-liquid-loading conditions, the realization of highly localized topological resonant modes and liquid-concentration sensing have not yet been reported. This gap has obviously limited the development of topological SAW sensors for micro-volume liquid samples.

Here, we construct valley-Hall topological phononic crystals on a lithium niobate substrate under a surface liquid loading. A straight waveguide is coupled to a resonant cavity to generate high-quality-factor topological SAW resonances under a liquid loading. Numerical simulations and experimental measurements are used to investigate the topological SAW resonance behaviors under a surface liquid loading. Besides, we further examine the responses of the device to concentration variations of NaCl and glucose aqueous solutions. The resonance frequency exhibits systematic shifts with the increasing concentration, and the experimental results agree well with the simulations. This work experimentally demonstrates a topologically protected SAW sensor for micro-volume liquid samples and provides a foundation for its potential application in biomedical sensings.

## 2. Model

The device is based on a topological PnC formed by a periodic array of etched holes in a 128° Y-cut $LiNbO_3$ substrate covered by a thin liquid loading, as illustrated in Fig. 1(a). The PnC comprises three types of unit cells, denoted Type-A, Type-B, and Type-C. All holes are etched to a depth of 4 μm but have different radii. The radius variation breaks spatial inversion symmetry and induces a topological phase transition. A liquid tank with a 15 μm height surrounds the PnC region to confine the liquid-loading area. Interdigital transducers (IDTs) are fabricated on both sides to excite and receive SAWs. Upon liquid loading, the holes are filled with the liquid, forming solid–liquid coupled acoustic-column structures that participate in the resonance process. Deionized water, with a sound speed of 1496 m/s and a density of 1000 kg/m$^3$, is initially used as the loading medium for band-structure analysis.

Finite-element simulations were performed to calculate the dispersion relations of PnC unit cells under liquid-layer loading. For the fluid–solid coupled system, COMSOL Multiphysics was used to couple the Solid Mechanics, Pressure Acoustics, and piezoelectric interfaces. Floquet periodic boundary conditions were applied to the unit cells for band-structure calculations. At the fluid–solid interface, a thermoviscous-acoustics impedance boundary condition was imposed.

Fig.1 presents the dispersion relations of liquid-loaded Type-A ( $r_A = 6$ μm, $r_B = 10$ μm ), Type-B ($r_A = r_B = 10$ μm), and Type-C ($r_A = 10$ μm, $r_B = 6$ μm) unit cells. Under liquid loading, bands associated with fluid–solid coupling are introduced within the SAW sound line. The dispersion relations in the absence of liquid loading are provided in **Appendix A**. As shown in Fig. 1(c), a closed Dirac cone is obtained at the K point of the first Brillouin zone when the two scatterers have identical radii (Type-B). In the Type-A and Type-C unit cells, spatial inversion symmetry is broken by the introduced radius asymmetry. Consequently, the valley degeneracy is lifted, a bandgap is opened, and the valley-Hall effect is induced.

To validate the finite-element results, the band structures were also calculated using a semi-analytical plane wave expansion–finite element method. The theoretical results are shown as blue dashed lines in Figs. 1(b–d) and are in good agreement with the numerical simulations. The insets in Figs. 1(b–d) show the displacement fields and energy flux at the K point. Displacement distributions are presented as colour maps, and energy-flow directions are indicated by red arrows. Opposite energy flow directions are observed for the Type-A and Type-C unit cells. This behavior is a key signature of valley modes and indicates the occurrence of a topological phase transition. Only part of the complete band structures is shown in Figs. 1(b–d), and the full band structures are provided in **Appendix A**. For the semi-analytical theoretical calculation of the plane wave expansion-finite element method, see **Appendix B**. When the unit-cell type is switched between Type-A and Type-C, a pronounced sign reversal of the Chern number is obtained, further confirming the topological phase transition. Details of the calculation are provided in **Appendix C**.

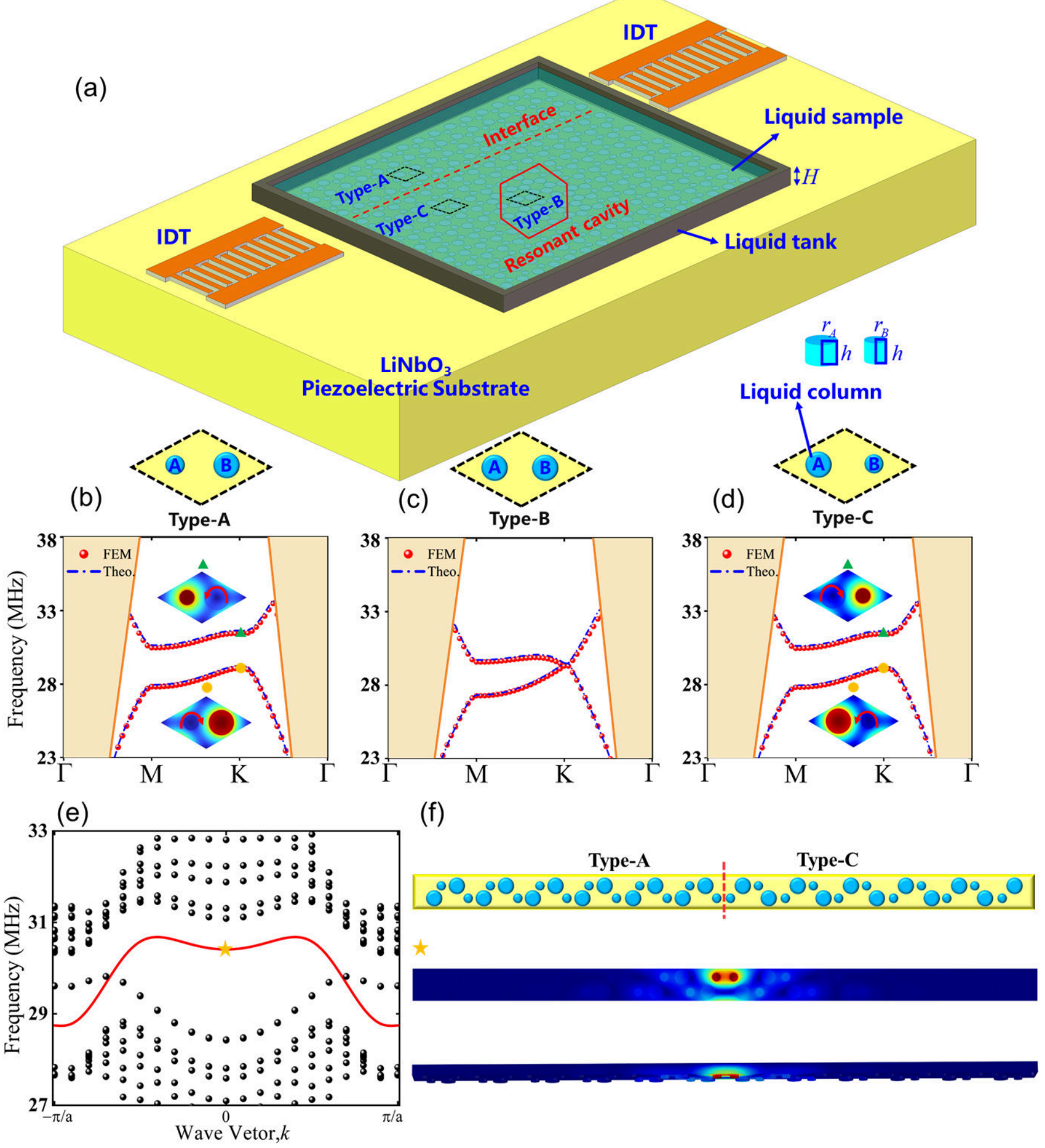

Fig. 1. Schematic of the topological SAW PnC device under surface liquid loading. (a) The PnC consists of a 128° Y-cut $LiNbO_3$ substrate and two sets of hole-based PnC scatterers patterned in the substrate. Three unit-cell configurations are denoted Type-A, Type-B, and Type-C. The lattice constant is $a = 24\ \mu m$. The PnC region is enclosed by a photoresist tank to confine the liquid, with a wall height of 15 μm. Interdigital transducers are positioned on both sides for SAW excitation and signal reception. Band structures of the (b) Type-A $(r_A = 6\ \mu m, r_B = 10\ \mu m)$, (c) Type-B $(r_A = r_B = 10\ \mu m)$, and (d) Type-C $(r_A = 10\ \mu m, r_B = 6\ \mu m)$ unit cells, respectively. A bandgap opens at the Dirac cones at the K and K'points of the first Brillouin zone for the Type-A and Type-C unit cells. Blue dashed lines indicate the theoretical results. Insets show the energy-flux distributions of PnCs with different topological phases. (e) Supercell band structure, in which black dots represent bulk states and the red line represents interface states. (f) Schematic of the supercell and energy-density distributions. The upper and lower panels show top and side views, respectively.

To assess energy localization, a supercell was constructed by using Type-A, B and C unit cells, and the topological interface states were calculated. The results are shown in Fig. 1(e). Black dots represent bulk states, whereas the red line denotes interface states located within the bandgap. The displacement field of the interface state marked by the star is shown in Fig. 1(f). The top and side views indicate that acoustic energy does not spread into the liquid bulk. Instead, it is strongly localized at the solid–liquid interfaces surrounding the holes. Most of the acoustic energy is confined near the substrate surface and within the holes, thereby reducing radiation loss into the liquid. This localization arises from the intrinsic confinement of the topological interface states. The hole structure also increases the solid–liquid contact area and may enhance sensing sensitivity. The effects of hole depth and liquid-loading thickness on the band structure, as well as the effects of device dimensions on the operating frequency, are provided in **Appendix D**.

## 3. Characteristics of the Coupled Topological Waveguide–Resonant Cavity System under a Surface Liquid Load

A defect resonant cavity was formed by replacing a local region of the Type-C PnC with Type-B symmetric unit cells, as shown in Fig. 2(a). The defect supports a mode within the bandgap, thereby localizing acoustic energy within a finite region and forming a resonant mode. Fig.2(c) shows the eigenfrequency spectrum of the resonant cavity. Blue dots represent the bulk states, whereas red dots denote cavity localized modes within the bandgap. The displacement field shown in Fig. 2(b) indicates strong energy confinement within the resonant cavity, providing a favorable basis for acoustic interactions with the liquid analyte.

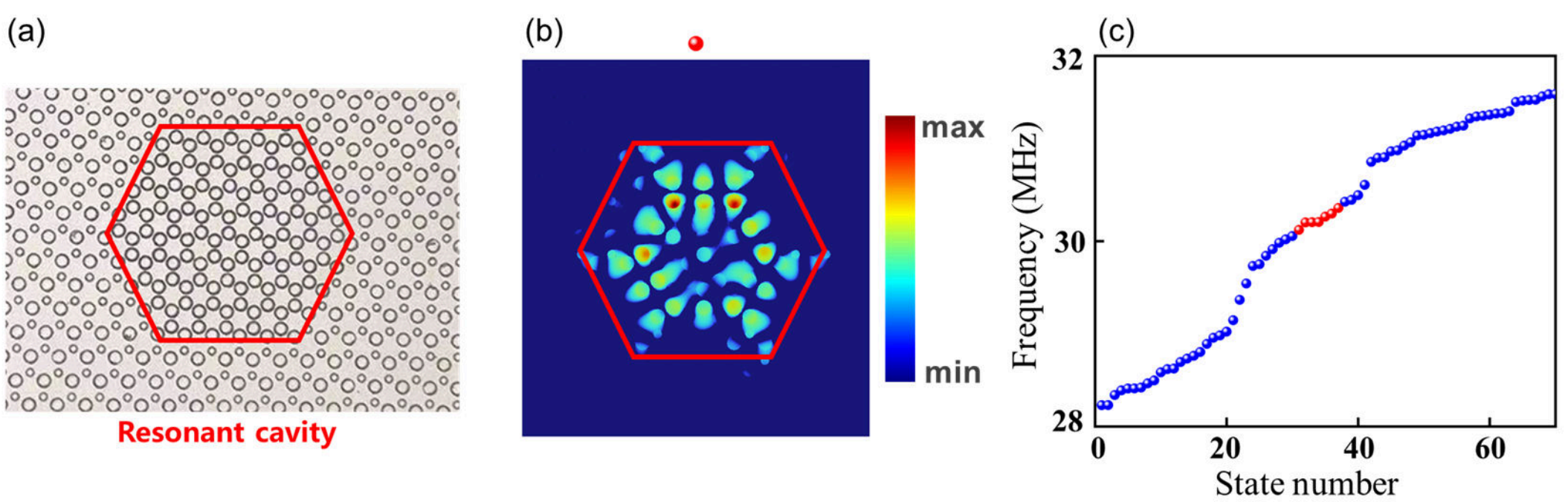

Fig. 2. Resonant cavity structure and localized modes. (a) Scanning electron micrograph of the resonant cavity. (b) Displacement field distribution of the resonant cavity. (c) Eigenfrequency spectrum. Blue dots represent bulk states, whereas red dots denote the cavity localized mode with concentrated energy shown in (b).

Based on the resonant cavity mode described above, a coupled waveguide–resonant cavity system was formed by placing a straight topological waveguide adjacent to the resonant cavity. A scanning electron micrograph of the device and a photograph of the packaged sample are shown in Fig. 3(a). A topological straight waveguide was formed along the interface between the Type-A and Type-C PnCs. The resonant cavity, composed of Type-B PnC unit cells, was introduced into the Type-C PnC region.

The device was fabricated using microfabrication processes, as detailed in **Appendix E**. The device has a footprint of 1.8 mm × 1.2 mm. A liquid tank was used to define the liquid-loading region. Interdigital transducers were positioned at both ends of the waveguide for excitation. Transmission measurements were performed after loading with deionized water. The input signal was supplied by a signal generator (Agilent E4428C, 250 kHz–3 GHz), and the SAW field distribution was acquired using a laser vibrometer (Polytec MSA 600).

When the frequency of the acoustic wave propagating in the waveguide matches the eigenfrequency of the resonant cavity, energy is coupled into the cavity and a resonance is formed. A pronounced resonant feature is observed in the transmission spectrum, as shown in Fig. 3(b). The green solid line represents the simulation result, whereas the blue dashed line represents the experimental result. Pronounced transmission dips are observed in both cases.

Figs. 3(c) and 3(d) show the displacement-field distributions at the waveguide passband frequency and the resonant frequency, respectively. Under passband conditions, acoustic waves propagate primarily along the waveguide interface, where the acoustic energy is concentrated and no appreciable scattering is observed. The experimentally measured transmission frequency was 29.51 MHz, corresponding to a deviation of 2.4% compared with a simulated value of 30.22 MHz. Under resonant conditions, the waveguide mode interferes with the resonant cavity mode, and the acoustic energy is strongly localized within the cavity. The corresponding transmission spectrum exhibits a resonant dip. The frequency discrepancy between experiment and simulation is primarily attributed to variations in liquid loading and ambient temperature.

Before fabrication, the effects of different coupling distances on device performance were calculated. A coupling distance of one lattice constant was selected.

The effects of sidewall taper angle on device performance were also evaluated. The results indicate that the operating frequency and Q factor are only weakly affected when the taper angle is sufficiently large. Further details are provided in **Appendix F.**

This coupling mechanism combines the robustness of topological transport with the high energy density of the resonant cavity. Strong energy localization enhances solid–liquid interactions and is expected to improve sensing sensitivity. Topological protection further helps maintain the linewidth and stability of the resonant response. Together, these characteristics provide a foundation for micro-volume liquid sample sensings.

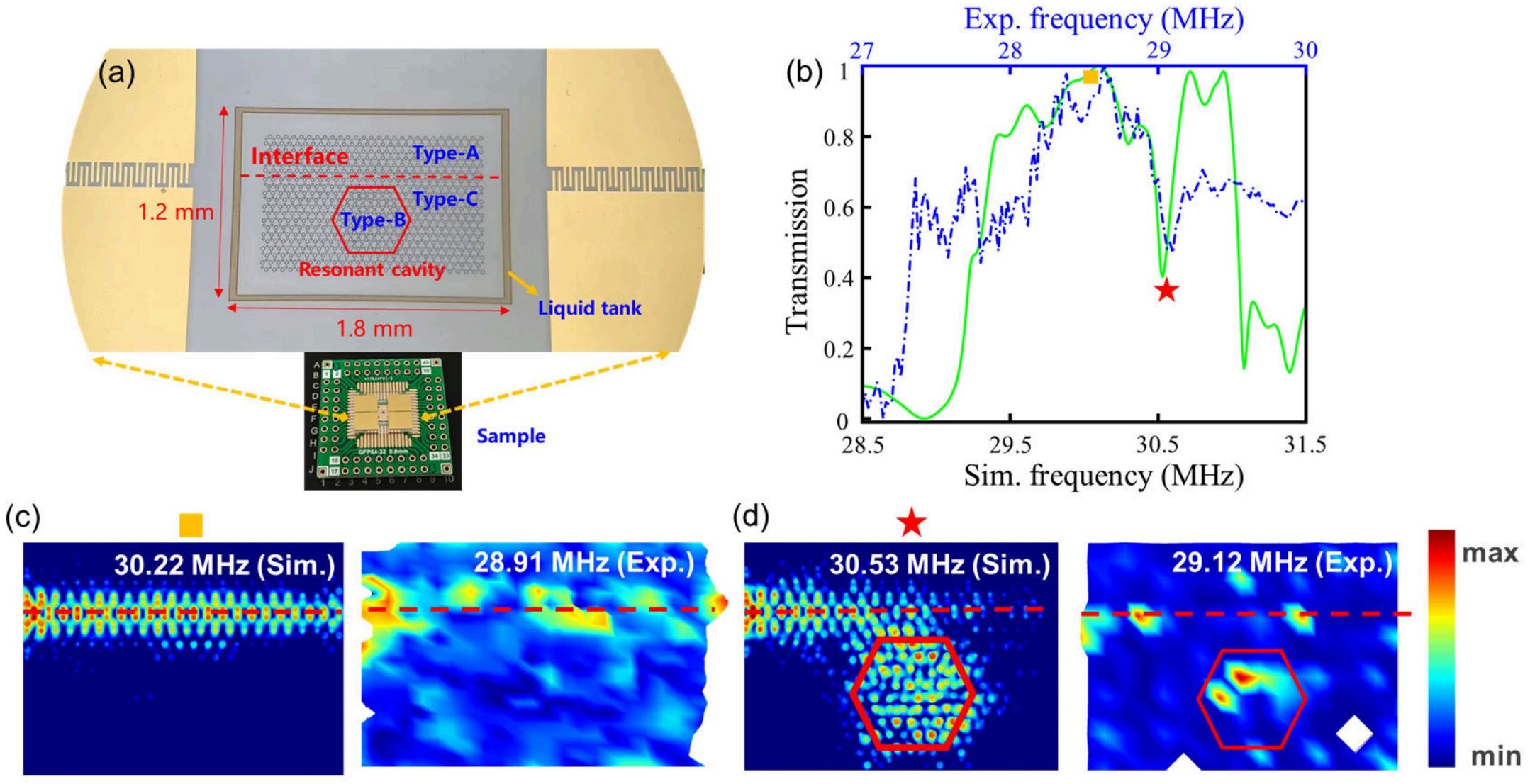

Fig. 3. Coupled waveguide–resonant cavity system and resonant characteristics under liquid loading. (a) Scanning electron micrograph and photograph of the coupled system comprising a topological interface waveguide and a resonant cavity. (b) Normalized transmission spectrum of the coupled system. The resonant mode and waveguide mode are marked by a red star and a yellow square, respectively. The green solid line and blue dashed line represent the simulated and experimental results, respectively. Simulated (c) and experimentally measured (d) displacement fields of the waveguide mode and resonant mode, respectively.

## 4. Experimental Validation of Sensing Performance

Based on the preceding analysis, experimental measurements and simulations were compared to validate the concentration-sensing performance of the device. Aqueous NaCl solutions were selected as the test analyte. The concentration range was 0–2%. Transmission spectra and modal field distributions were obtained after sequential loadings of solutions at different concentrations.

Fig. 4(a) shows the transmission spectra obtained at different NaCl concentrations. The resonant frequency shifts towards higher frequencies as the concentration increases. The extracted resonant frequencies are fitted in Fig. 4(b). In the simulations, the resonant frequency shifts from 30.53 to 30.95 MHz as the concentration increases from 0% to 2%, yielding an average sensitivity of 210 kHz/%. Experimentally, the resonant frequency shifts from 29.12 to 29.50 MHz, corresponding to an average sensitivity of 190 kHz/%. The frequency deviation between the simulated and experimental results is 4.84%. This discrepancy is primarily attributed to small variations in liquid-layer thickness and to ambient-temperature fluctuations, which affect the sound velocity and elastic constants of the substrate. Overall, good agreement is obtained between the simulated and experimental results, supporting the reliability of the sensing responses.

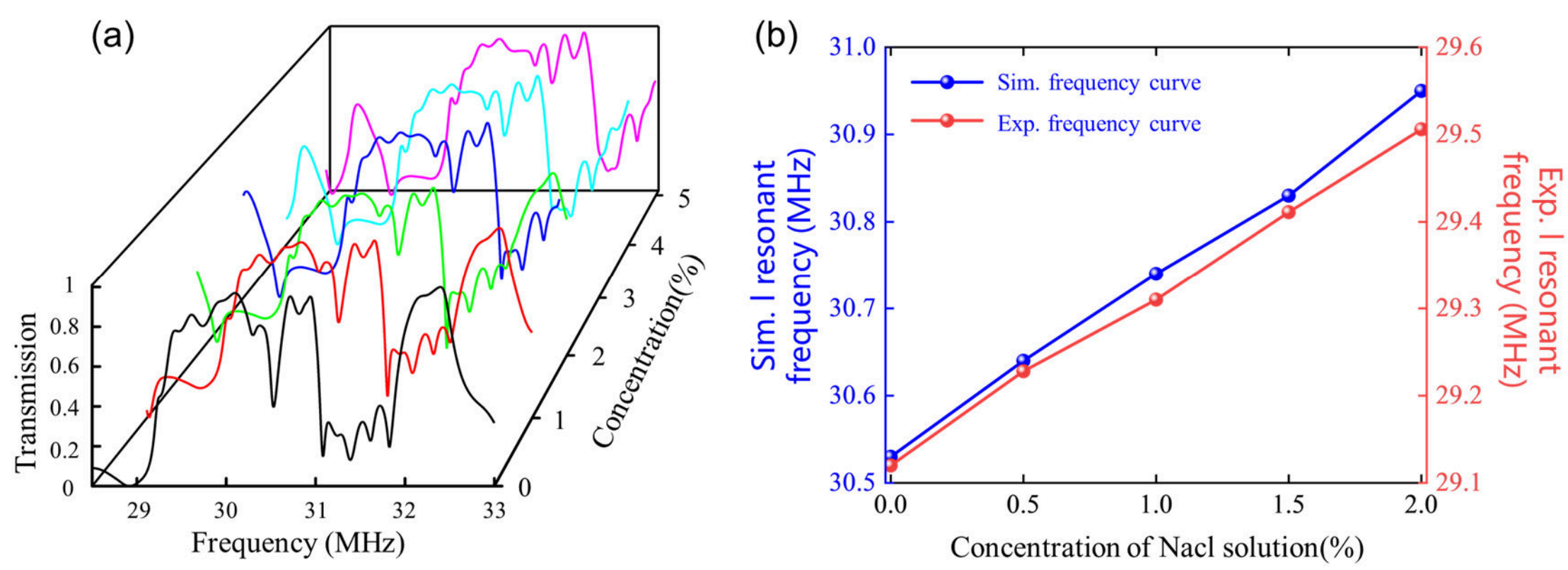

Fig. 4. Sensing characteristics of the topological SAW coupled device for NaCl concentrations in aqueous solution. (a) Transmission spectra acquired at NaCl concentrations ranging from 0% to 5%. (b) Simulated and experimentally measured resonant frequencies as a function of NaCl concentration from 0% to 2%.

To further assess the general sensing applicability of the device, glucose aqueous solutions were examined under the same sensing conditions over a concentration range of 1–10%. As shown in Fig. 5, the resonant frequency also shifts towards higher frequencies with increasing glucose concentration. The simulated average sensitivity is 109.4 kHz/%, whereas the experimentally measured average sensitivity is 101.8 kHz/%. The difference between the two values is 7.5%, indicating good agreements.

Both experiments indicate that an increase in liquid concentration produces an increase in resonant frequency. Stable sensing responses were obtained for both solutes, suggesting potential applications in a range of biochemical sensing scenarios. In addition, only 0.04 μL of sample is required for a single measurement. This volume is substantially lower than that typically required for conventional SAW sensors and is suitable for micro-volume biological-sample analysis.

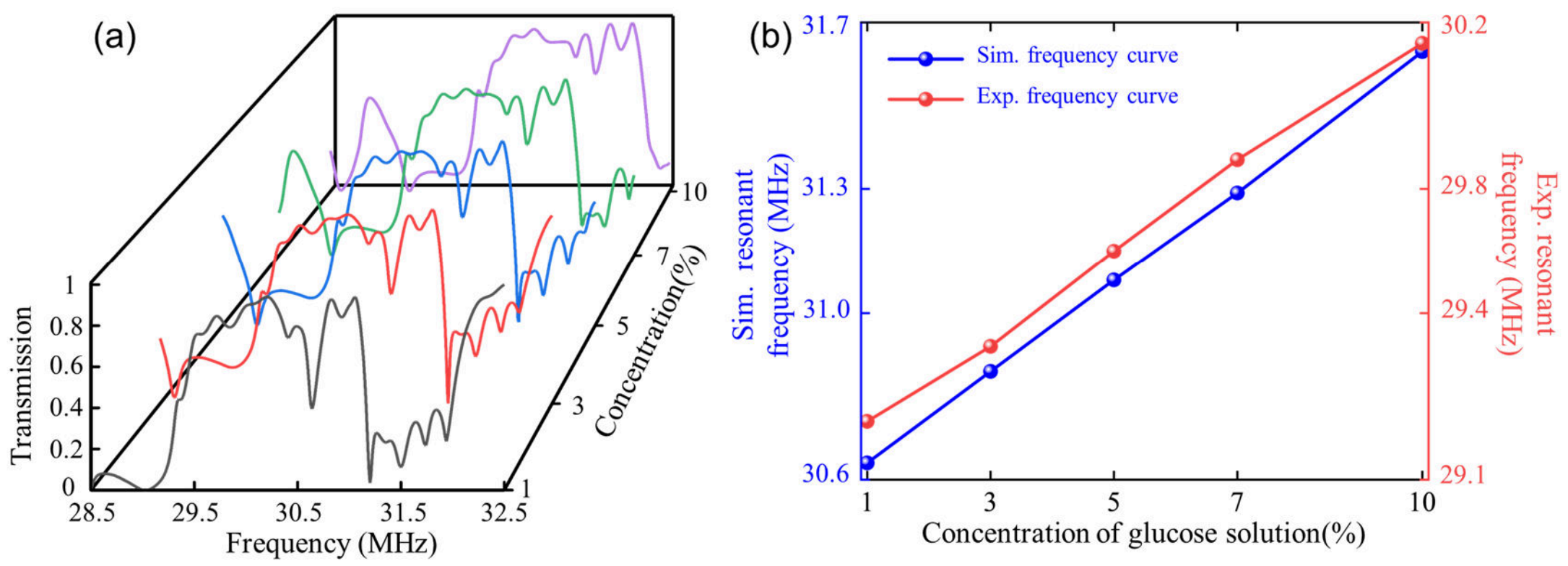

Fig. 5. Sensing characteristics of the topological SAW resonant device for glucose concentration. (a) Transmission spectra acquired at glucose concentrations ranging from 1% to 10%. (b) Simulated and experimentally measured resonant frequencies as a function of glucose concentration.

The resolution of the proposed topological SAW sensor was further evaluated using several low-concentration NaCl and glucose solutions. The results are summarized in Table 1. For a small concentration variation of 0.01%, frequency shifts of 2 kHz and 1 kHz were resolved for NaCl and glucose solutions, respectively, indicating high frequency resolution of the sensor.

Table 1 Frequency resolution of the device for low-concentration NaCl and glucose solutions

| Concentration | 0.01% | 0.02% | Frequency shift | 0.1% | 0.2% | Frequency shift |
|---|---|---|---|---|---|---|
| NaCl Solution | 30.523 (MHz) | 30.525 (MHz) | 2 kHz | 30.542 (MHz) | 30.563 (MHz) | 21 kHz |
| Glucose solution | 30.522 (MHz) | 30.523 (MHz) | 1 kHz | 30.532 (MHz) | 30.543 (MHz) | 11 kHz |

## 5. Conclusion

In this work, a topological resonant SAW sensor for micro-volume liquid sensings was designed and experimentally realized by using a topological waveguide–resonant-cavity system under a surface liquid loading. Under liquid loading, the device supports a topologically protected resonant mode with strong cavity confinement. The band structure was evaluated using a semi-analytical plane wave expansion–finite element method, and both waveguide transmission and cavity localization were verified experimentally. The sensor can operates with a sample volume of only 0.04 μL. Sensitivities of 190 kHz/% for NaCl and 101.8 kHz/% for glucose were obtained, in agreement with the simulated results. Besides, high resolutions were also achieved for solution concentration variations. This work provides an experimental basis for the development of stable, micro-volume sample SAW biosensing chips.

## Appendix A. Effect of a Water Loading on the Band Structure of Pnc

The band structure of Pnc shown in Fig.1 in the absence of liquid loading was calculated and is shown in Fig. S1. Without liquid loading, the two target bands are absent.

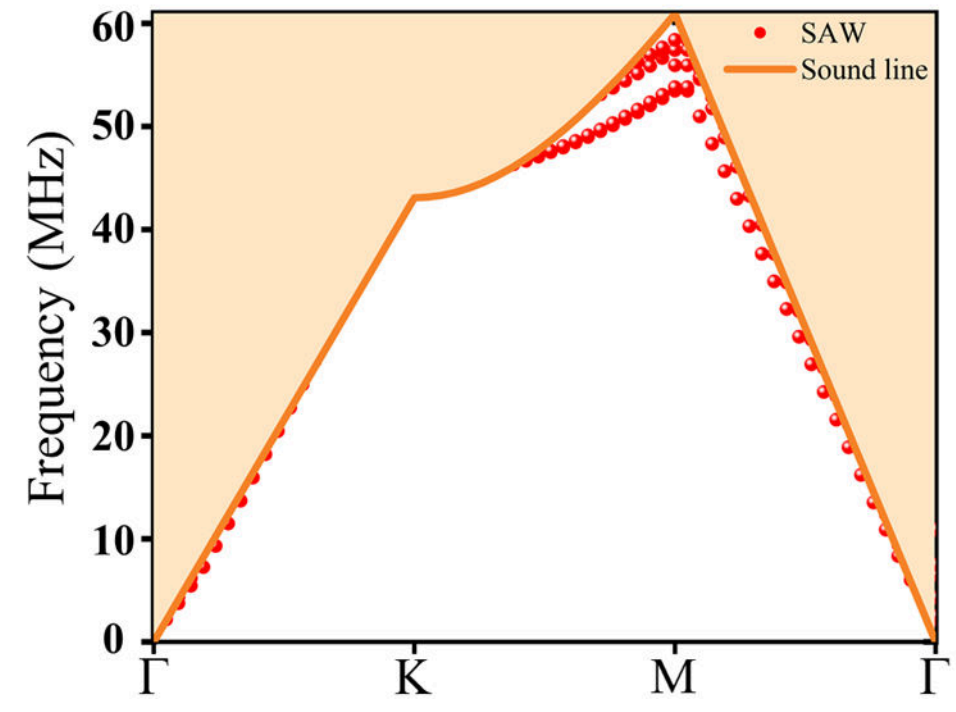

Fig.S1. Band structure without liquid loading.

The complete band structures of Pnc under a surface water loading is shown in Fig. S2.

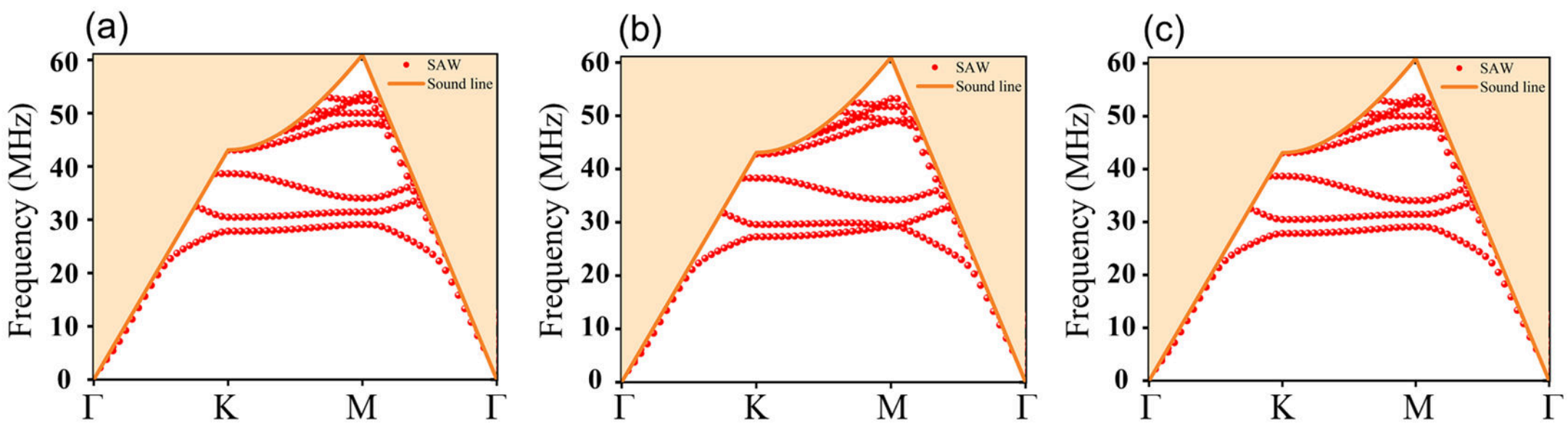

Fig.S2 Complete band structures of Pnc under a surface liquid loading.

## Appendix B. Plane-wave expansion-finite element semi-analytical method

We consider a class of fluid-loaded elastic structures that exhibit two-dimensional periodicity in the *x*-*y* plane and have a finite thickness along the z direction. The solid phase is assumed to undergo small deformations. The fluid phase is homogeneous, compressible, inviscid and irrotational, and effects of mean flow, gravity and thermal dissipation are neglected. The model is formulated from the governing equations of continuum mechanics via Bloch plane-wave expansion and Galerkin finite-element discretization.

The kinematic relations, constitutive equations and linear momentum balance for the solid are given by

$$\varepsilon_{ij}^{s}=\frac{1}{2}\left(u_{i,j}+u_{j,i}\right),\quad \sigma_{ij}^{s}=C_{ijkl}^{s}\varepsilon_{kl}^{s},\quad \sigma_{ij,j}^{s}+\omega^{2}\rho_{s}u_{i}=0, \tag{1}$$

where, u, $\varepsilon_{ij}^{s}$ and $\sigma_{ij}^{s}$ denote the solid displacement, infinitesimal strain and Cauchy stress, respectively; $\rho_s$ is the solid mass density and $C_{ijkl}^{s}$ is the anisotropic elastic stiffness tensor.

Under the linear acoustic approximation, the pressure governing equation for inhomogeneous acoustic media reads：

$$\nabla\cdot\left(\frac{1}{\rho_f}\nabla p\right)+\frac{\omega^{2}}{\rho_f c_f^{2}}p=0, \tag{2}$$

where，$p$ represents the acoustic pressure; $\rho_f$ and $c_f$ are the mass density and sound speed of the fluid, respectively.

Let **r**=(x,y) denote the in-plane position vector, k is the Bloch wave vector within the first Brillouin zone, and G is a two-dimensional reciprocal lattice vector. By Bloch's theorem, both the displacement and the acoustic pressure admit the expansion:

$$\psi(\mathbf{r},z)=\sum_{\mathbf{G}}\psi_{\mathbf{G}}(z)e^{i(\mathbf{k}+\mathbf{G})\cdot\mathbf{r}},\quad \psi\in(u_1,u_2,u_3,p) \tag{3}$$

Substituting Eq. (3) into the weak form of the solid equations and exploiting the orthogonality of the in-plane Fourier basis over the unit cell yields the elemental stiffness and mass matrices of the solid:

$$\mathbf{K}^{e}_{s,\mathbf{GG}'}=\int_{z_e}\mathbf{B}^{sH}_{\mathbf{G}}C^{s}_{\mathbf{G}-\mathbf{G}'}(z)\mathbf{B}^{s}_{\mathbf{G}'}dz,\quad \mathbf{M}^{e}_{s,\mathbf{GG}'}=\int_{z_e}\mathbf{N}^{uH}_{\mathbf{G}}\rho_{s,\mathbf{G}-\mathbf{G}'}(z)\mathbf{N}^{u}_{\mathbf{G}'}dz \tag{4}$$

Here, $\mathbf{K}^{e}_{s,\mathbf{GG}'}$ and $\mathbf{M}^{e}_{s,\mathbf{GG}'}$ are the solid stiffness and mass matrices, respectively; $\mathbf{B}^{sH}_{\mathbf{G}}$ is the strain-displacement matrix associated with the G-th plane-wave harmonic, and $C^{s}_{\mathbf{G}-\mathbf{G}'}(z)$ denotes the Fourier convolution coefficient of the solid elastic stiffness tensor.

The same projection is applied to the acoustic domain. Let $(\chi_f(\mathbf{r},z))$ be the fluid indicator function, taking the value unity in the internal pores and the surface liquid layer and zero within the solid domain. The acoustic stiffness and mass matrices then read:

$$\mathbf{K}^{e}_{a,\mathbf{GG}'}=\int_{z_e}\mathbf{D}^{H}_{\mathbf{G}}\left(\frac{\chi_f}{\rho_f}\right)_{\mathbf{G}-\mathbf{G}'}\mathbf{D}_{\mathbf{G}'}dz,\mathbf{M}^{e}_{a,\mathbf{GG}'}=\int_{z_e}\mathbf{N}^{H}_{p}\left(\frac{\chi_f}{\rho_f c_f^2}\right)_{\mathbf{G}-\mathbf{G}'}\mathbf{N}_{p}dz, \tag{5}$$

where, $\mathbf{K}^{e}_{a,\mathbf{GG}'}$ and $\mathbf{M}^{e}_{a,\mathbf{GG}'}$ denote the acoustic stiffness and mass matrices, $\mathbf{D}^{H}_{\mathbf{G}}$ is the pressure-gradient operator matrix for the G-th plane-wave harmonic, and $\left(\frac{\chi_f}{\rho_f}\right)_{\mathbf{G}-\mathbf{G}'}$ is the Fourier convolution coefficient of the fluid parameter $(\frac{\chi_f}{\rho_f})$.

The traction exerted on the solid by the fluid is governed solely by the normal component of the acoustic pressure:

$$\boldsymbol{\sigma}_s\cdot\mathbf{n}_s=-p\mathbf{n}_s. \tag{6}$$

Normal velocity continuity holds across the fluid-solid interface. This kinematic constraint can be recast as a boundary condition on the normal derivative of pressure:

$$\mathbf{n}_f\cdot\mathbf{v}=-i\omega\mathbf{n}_f\cdot\mathbf{u},\quad \frac{1}{\rho_f}\frac{\partial p}{\partial n_f}=\omega^2\mathbf{n}_f\cdot\mathbf{u} \tag{7}$$

Assembling all foregoing relations yields the discretized governing equations for the solid deformation and acoustic pressure fields:

$$\mathbf{K}_s(\mathbf{k})\mathbf{d} + \mathbf{B}_{up}\mathbf{p} = \omega^2\mathbf{M}_s\mathbf{d} \tag{8}$$

$$\mathbf{K}_a(\mathbf{k})\mathbf{p} = \omega^2[\mathbf{M}_a\mathbf{p} - \mathbf{B}_{up}^H\mathbf{d}] \tag{9}$$

Combining Eqs. (8) and (9) leads to the generalized eigenvalue problem of the coupled system:

$$\begin{bmatrix} \mathbf{K}_s(\mathbf{k}) & \mathbf{B}_{up} \\ 0 & \mathbf{K}_a(\mathbf{k}) \end{bmatrix} \begin{bmatrix} \mathbf{d} \\ \mathbf{p} \end{bmatrix} = \omega^2 \begin{bmatrix} \mathbf{M}_s & 0 \\ -\mathbf{B}_{up}^H & \mathbf{M}_a \end{bmatrix} \begin{bmatrix} \mathbf{d} \\ \mathbf{p} \end{bmatrix}. \tag{10}$$

The band frequencies of the periodic structure are retrieved by solving for the nontrivial solutions of the resulting matrix equation.

## Appendix C. Calculation of the Topological Phase Transition

Fig. S3(a) shows the variation in bandgap as a function of the radius difference between the scatterers. A topological phase transition occurs when the radius difference is zero. To verify this transition, the valley Chern number and Berry curvature were calculated. Both parameters characterize the topological properties of the dispersion bands. Berry curvature and the valley Chern number were quantitatively evaluated using a discrete method. The nontrivial topology of the SAW system is indicated by a non-zero valley Chern number, which is defined as the integral of the Berry curvature around the K and K' valleys:

$$\boldsymbol{C}_{\mathrm{V}}^{(n)} = \frac{1}{2\pi}\int \boldsymbol{\Omega}(\mathbf{k})d^2\mathbf{k}.$$

Here, Ω(k) denotes the Berry curvature:

$$\boldsymbol{\Omega}(\mathbf{k}) = \mathrm{i}\nabla_{\mathrm{k}} \times \langle \mathbf{u}(\mathbf{k}) | \nabla_k | \mathbf{u}(\mathbf{k}) \rangle.$$

The eigenmode displacement field, u(k), was obtained from COMSOL Multiphysics simulations. The Berry-curvature distributions are shown in Fig. S3(b, c) and are primarily localized near the K and K' valleys. When the unit-cell configuration is switched between Type-A and Type-C, a pronounced sign reversal is observed, confirming the topological phase transition. The calculated valley Chern number is smaller than 1/2, which is attributed to strong spatial-inversion-symmetry breaking[19–21]. A large radius contrast between the Type-A and Type-C scatterers produces strong inversion-symmetry breaking. This weakens the localization of the Berry curvature near the valleys, causes the valley Chern number to deviate from its theoretical value, and enhances intervalley coupling.

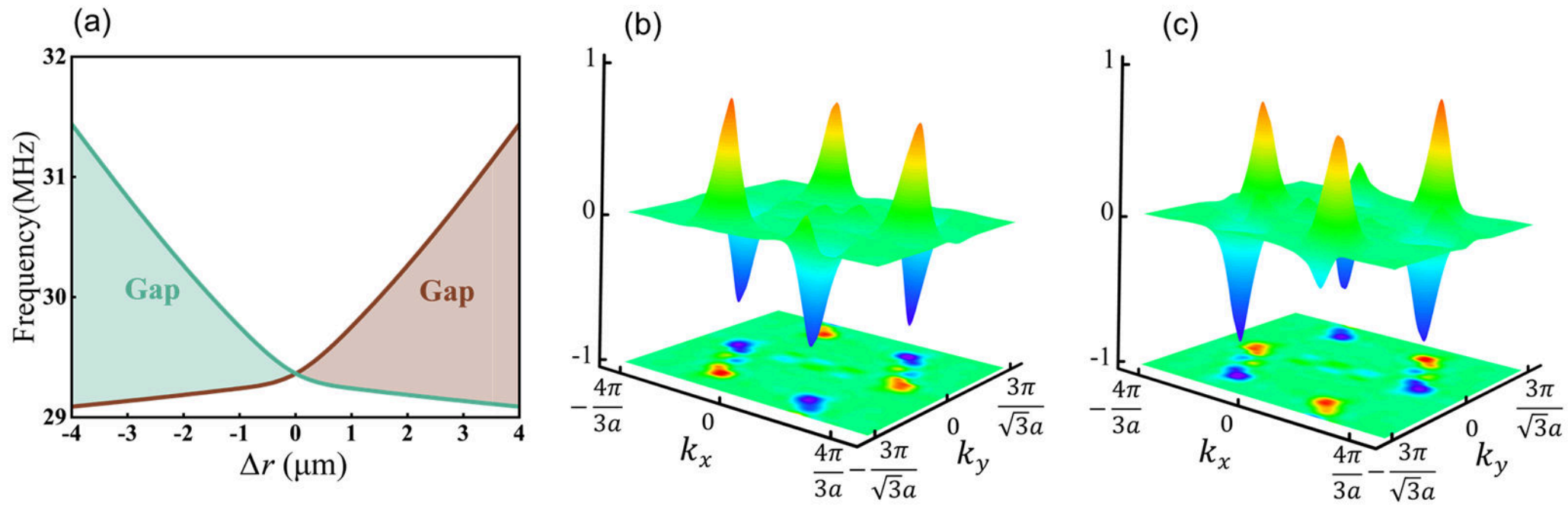

Fig. S3. Bandgap evolution and topological phase transition of liquid-loaded PnCs. (a) Dependence of the bandgap on the radius difference between the scatterers in the unit cell. (b, c) Berry-curvature distributions around the K and K' points, respectively.

## Appendix D. Effects of Design Parameters on the Operating Frequency

In the proposed structure, a phononic bandgap is formed by etching holes into the $LiNbO_3$ surface and applying a liquid-loading layer. The effects of etching depth and liquid-loading thickness on the band structure were investigated. As shown in Fig. S4(a), the bandgap increases with increasing etching depth (water-column height). Beyond 4 μm, the bandgap width approaches saturation, whereas the operating frequency continues to decrease. The liquid-loading thickness also markedly affects the operating frequency, as shown in Fig. S4(b). This parameter provides a convenient means of tuning the operating frequency over different ranges. Fig.S4(c) shows the interface-state band structure of the supercell.

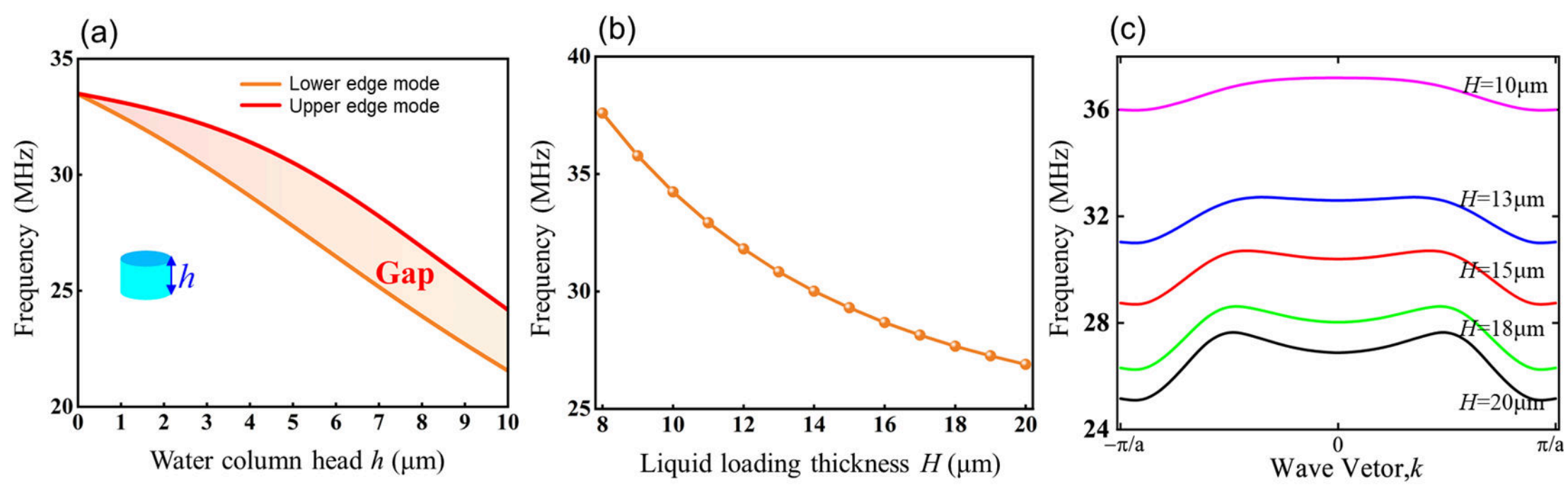

Fig. S4. Simulated dependence of the PnC bandgap and operating frequency on etching depth (a) and liquid-loading thickness (b). The inset in (a) illustrates the etching depth (water-column height). The red and orange curves represent the upper-edge and lower-edge frequencies of the bandgap,

respectively. (c) The interface bands of different liquid loading thickness.

The effect of device dimensions on the operating frequency was further examined, as shown in Fig. S5. When all geometrical parameters are scaled by the same factor, the operating frequency changes substantially. This result indicates that the operating frequency can be tuned through device scaling for different application scenarios.

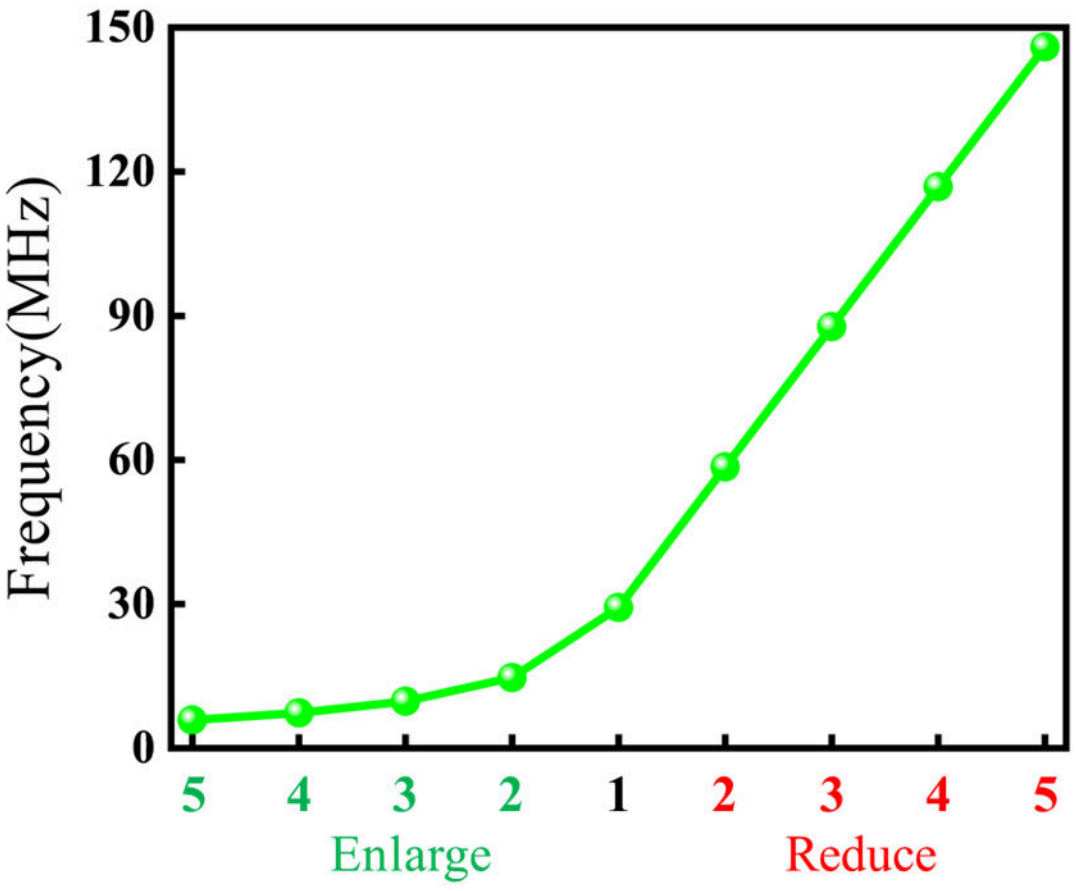

Fig. S5. Tuning of the operating frequency using different geometric scaling factors. The horizontal axis denotes the overall enlargement or reduction factor.

## Appendix E. Fabrication Process

The device was fabricated on a 128° rotated Y-cut lithium niobate substrate using high-precision microfabrication processes. Fabrication consisted of two main stages: topological PnC patterning and split-finger interdigital-transducer fabrication. The complete fabrication process is shown in Fig. S8.

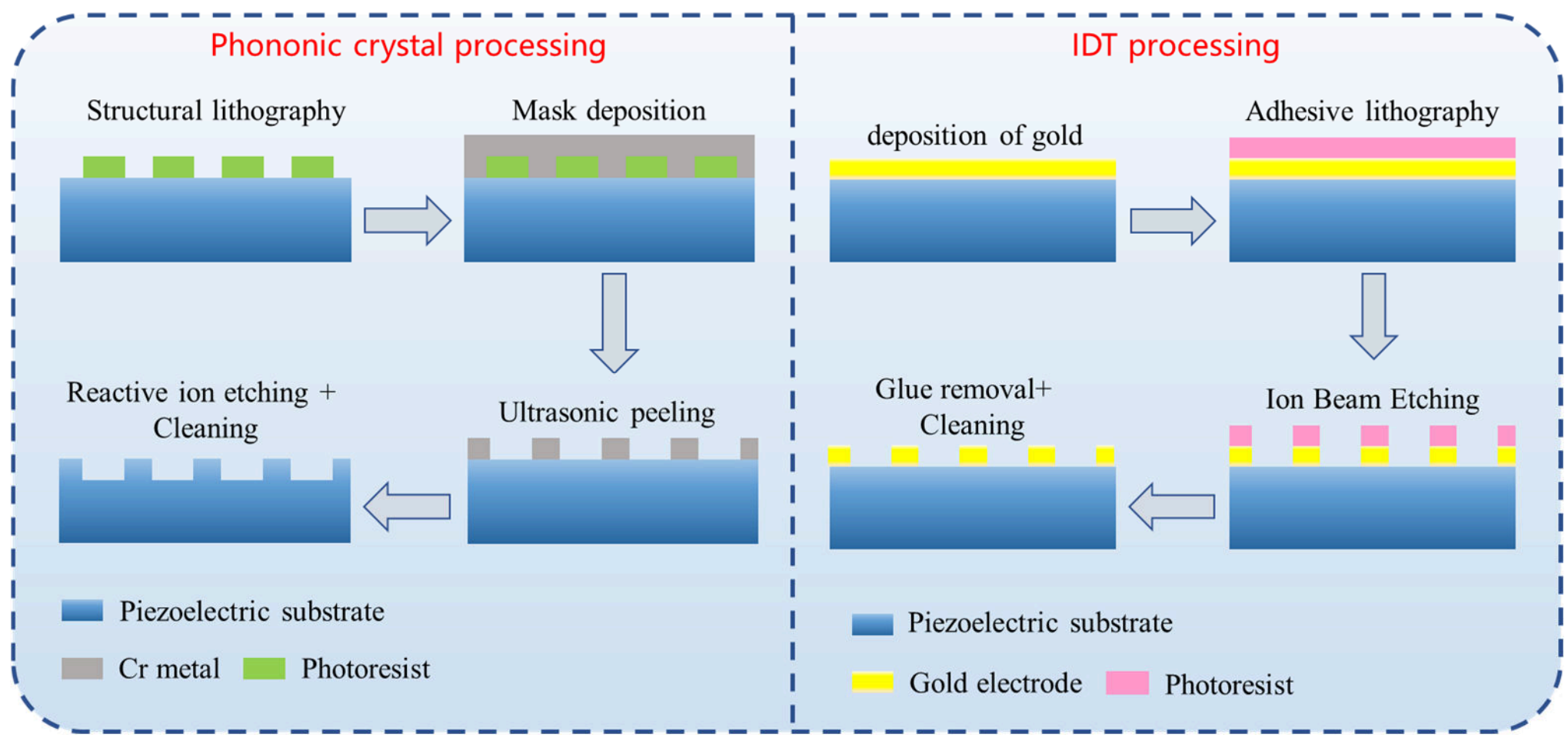

Fig.S8. Fabrication process for the PnC structure and gold interdigital transducers.

## Appendix F. Effects of Design Parameters on the Quality Factor

The effects of coupling distance on the coupled topological waveguide–resonant cavity system was calculated, as shown in Fig. S6. The coupling distance strongly affects the Q factor. When the distance increases to 2a, Q factor decreases substantially. At a distance of 3a, the resonant response is no longer clearly discernible. The displacement fields show the same trend. The strongest waveguide-cavity interaction and the highest Q factor are obtained at the smallest coupling distance. As the distance increases, the cavity exerts a weaker influence on the topological waveguide and the resonant response gradually diminishes. The Q factor was extracted using the conventional 3 dB bandwidth method.

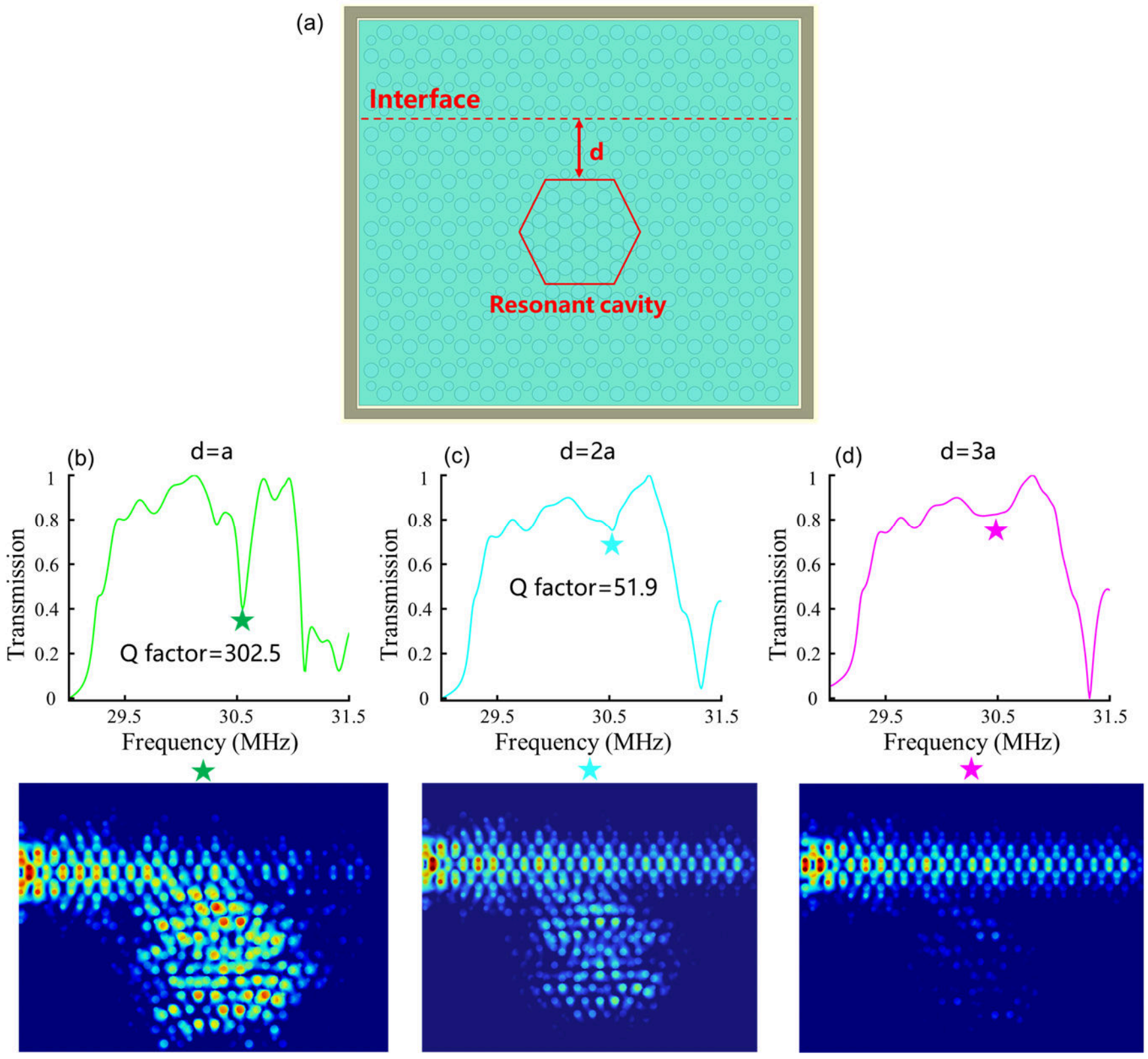

Fig. S6. Effect of coupling distance on the device quality factor. (a) Schematic of the coupled device, where $d$ is the distance between the topological waveguide and resonant cavity. (b–d) Transmission spectra and resonant displacement fields for coupling distances of $d = a$, $2a$, and $3a$, respectively. Here, $a$ denotes the lattice constant.

During fabrication, the holes may exhibit a small sidewall taper. This non-ideal geometric feature was included in the simulations to assess its influence on device performance. The effects of side wall angle on the bandgap and Q factor are shown in

Fig. S7. The bandgap is only weakly affected by the taper angle. In contrast, the Q factor is strongly affected at small angles and becomes relatively stable when the angle exceeds 80°. A side wall angle of 90° was therefore selected for fabrication. Minor deviations from this angle are expected to have limited effects on device performance.

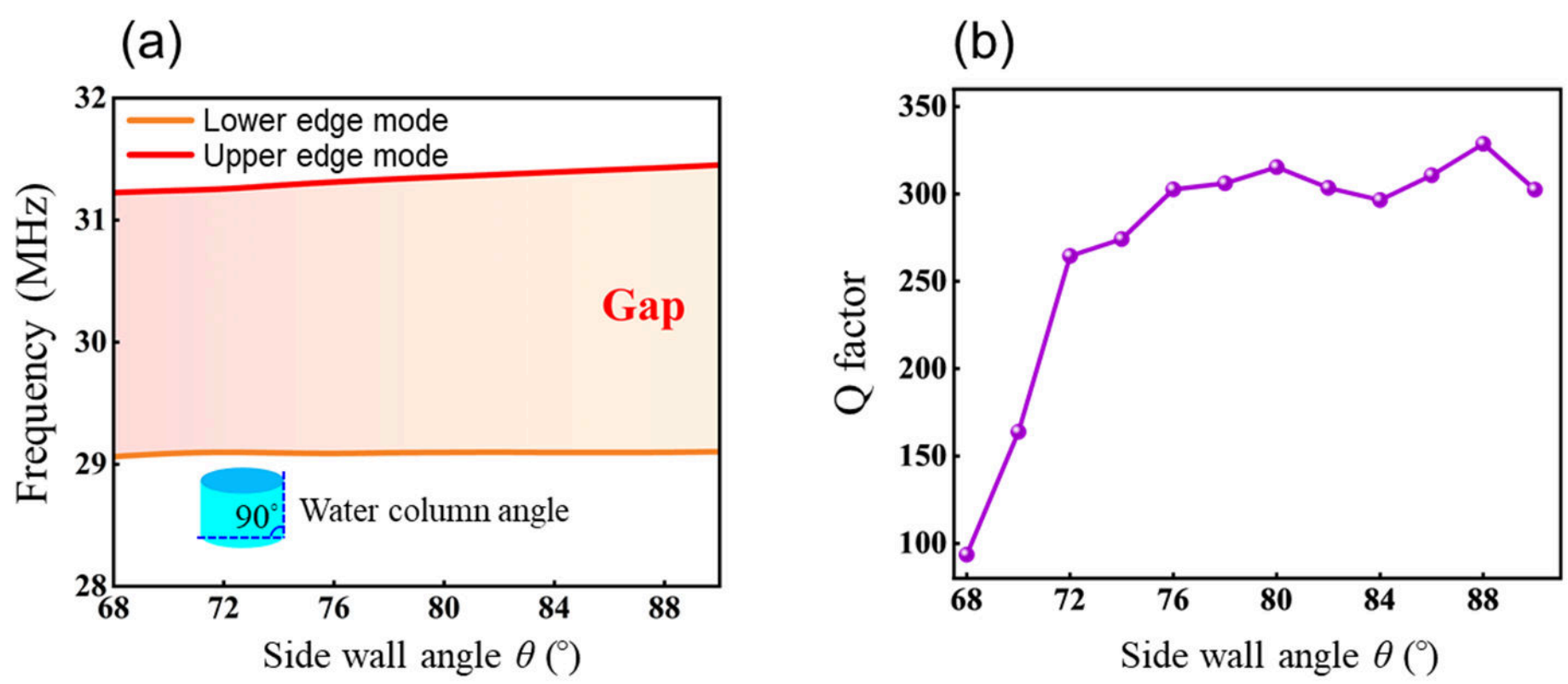

Fig. S7. Effects of hole side wall angle on the operating bandgap (a) and the Q factor(b). The angle has a limited effect on the operating frequency but strongly affects the Q factor.